\documentclass[letterpaper, 9 pt, conference]{ieeeconf}  

\IEEEoverridecommandlockouts                              

\usepackage{gensymb}
\usepackage{multicol}
\usepackage{enumerate}
\usepackage{graphicx}

\usepackage{balance}
\usepackage{subcaption}
\usepackage{listings}
\usepackage[T1]{fontenc}
\usepackage{dblfloatfix}

\usepackage{eso-pic}
\newcommand\AtPageUpperMyright[1]{\AtPageUpperLeft{
 \put(\LenToUnit{0.5\paperwidth},\LenToUnit{-1cm}){
     \parbox{0.5\textwidth}{\raggedleft\fontsize{9}{11}\selectfont #1}}
 }}
\newcommand{\conf}[1]{
\AddToShipoutPictureBG*{
\AtPageUpperMyright{#1}
}
}

\title{\Huge
TrappeD: DRAM Trojan Designs for Information Leakage and Fault Injection Attacks\\ \vspace{2mm} 
 \vspace{-4mm}
 }
 
\conf{20th International Workshop on
Microprocessor/SoC Test, Security \& Verification (MTV19), December 2019}

\author{%
Karthikeyan Nagarajan,  Asmit De, Mohammad Nasim Imtiaz Khan, and Swaroop Ghosh\\
School of EECS, The Pennsylvania State University, Univeristy Park, USA\\ 
 $\{$kxn287, asmit, muk392, szg212$\}$@psu.edu 
\vspace{-5mm}
}

\begin{document}
\maketitle
\thispagestyle{empty}
\pagestyle{empty}

\begin{abstract}
In this paper, we investigate the advanced circuit features such as  wordline- (WL) underdrive (prevents retention failure) and overdrive (assists write) employed in the peripherals of Dynamic RAM (DRAM) memories from a security perspective. In an ideal environment, these features ensure fast and reliable read and write operations. However, an adversary can re-purpose them by inserting Trojans to deliver malicious payloads such as, fault injections, Denial-of-Service (DoS), and information leakage attacks when activated by the adversary. Simulation results indicate that wordline voltage can be increased to cause retention failure and thereby launch a DoS attack in DRAM memory. Furthermore, two wordlines or bitlines can be shorted to leak information or inject faults by exploiting the DRAM's refresh operation. We demonstrate an information leakage system exploit by implementing TrappeD on RocketChip SoC.
\end{abstract}


\section{Introduction}

Integrated Circuit (IC) fabrication has become increasingly vulnerable to malicious modifications in the form of Hardware Trojans \cite{Trojan_war} due to the outsourcing of semiconductor design and fabrication to third party fabrication houses. Ideally, these modifications need to be detected during pre-Silicon verification and  post-Silicon testing. But, it is possible to design these Trojans to remain dormant during the test phase and only activate under rare conditions. Once activated, the Trojans perform undesirable operations such as write/retention failures or even leak sensitive data. This threat is of special concern to government agencies, military \cite{DARPA}, technology developers, finance, and energy sectors. Recent news regarding the tampering of server motherboards by Chinese manufacturers that affected top US companies like Amazon, Apple etc. \cite{Amazon} provides as a strong motivation to investigate the possibility of hidden components in each step of the design and manufacturing process.

Memory Trojan can lead to read/write/retention failures and information leakage. In prior works, authors have proposed memory Trojan Trigger circuits and payloads that can evade testing phase and cause different failures. A Trojan for embedded SRAM is proposed in \cite{Hoque_HT_Embedded}. The authors use unique data patterns written to pre-selected address to trigger their Trojan. Note that these unique patterns are not tested during standard post-manufacturing memory tests and thereby remains undetected. That data pattern feeds the input of the Trojan payload transistors which short the data node of a victim SRAM cell to ground and corrupts the data. The feasibility of this Trojan \cite{Hoque_HT_Embedded} is limited since the payloads require tapping the bitcells which might not be possible since they are very compact. \cite{ENTT_Nagarajan} introduces an NVM-based Trojan trigger that leverages the non-volatility of Resistive RAM (RRAM). It possesses unique characteristics e.g., non-volatility and gradual drift in resistance with pulsing voltage. The triggers presented exploits RRAM's gradual resistance drift under pulsing current and its non-volatility to ensure that the hammering need not be consecutive. This allows the trigger to evade system-level detection techniques. 

\begin{figure} [b] 
 \vspace{-2mm}
 \begin{center}
    \includegraphics[width=.48\textwidth]{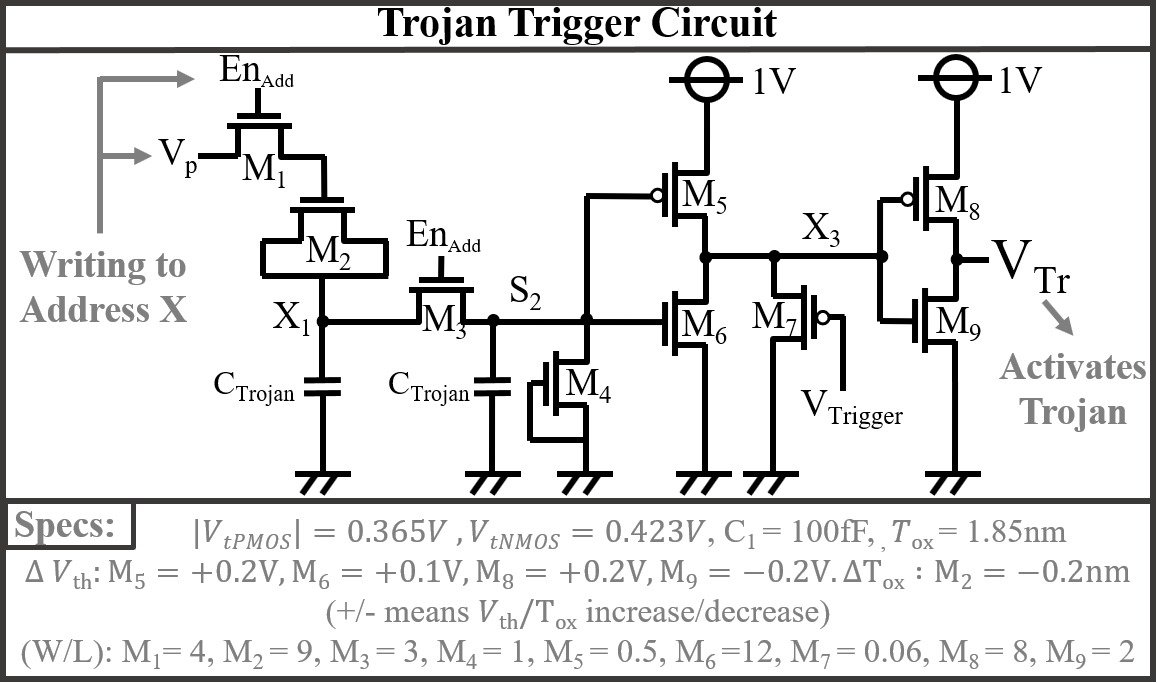}
 \end{center}
 \vspace{-4mm}
 \caption{Capacitor-based Trojan trigger circuit proposed in \cite{DATE_Nasim_2019}.} 
 \label{Trigger_DATE}
  \vspace{-1mm}
 \end{figure}

Designing a small, controllable and undetectable Trojan is the key element to deploy an efficient one. In \cite{A2_paper}, a capacitor-based analog Trojan trigger, A2, is presented which is controllable, stealthy and small. In \cite{DATE_Nasim_2019}, another capacitor-based Trojan trigger is proposed which gets activated if a pre-selected address is written with specific data patterns for a specific number of times. The proposed trigger circuit is shown in Fig. \ref{Trigger_DATE}. The circuit has two inputs, $En\textsubscript{Add}$ (the address enable signal of a pre-selected address) and $V\textsubscript{p}$ (a logic circuit generates $V\textsubscript{p}$ based on the data pattern). The work also presents payloads of the Trojan such as fault injection, Denial of Service (DoS) and information leakage attacks on emerging Non-Volatile Memories (NVMs). The advantage of such trigger lies in the fact that it can be hidden in the filler areas of the non-memory logic (e.g., address pre-decoding and pipelining units, also called midlogic) \cite{Intel_midlogic}. However, the payload of this Trojan also requires tapping the bitcell.

\textbf{Proposed Attack Model: } In this work, we assume an untrusted manufacturing house located outside the US that can alter the chip GDS-II file to introduce the malicious Trojan trigger and payload. This assumption is widely accepted in the hardware security community because of large filler areas present in the chip and the adversary's access to the raw design. We propose the use of a capacitor-based hardware Trojan trigger \cite{DATE_Nasim_2019} and novel payload circuits and perform detailed analysis to guarantee that the Trojan is, (i) triggered even under worst-case process and temperature conditions with correct inputs; (ii) able to bypass conventional post-manufacturing test. The Trojan is activated if a particular pre-selected address of L1 Cache is accessed for $\sim$1800 times. Note that the proposed Trojan trigger directly taps the wordline of the pre-selected address to leverage the existing decoder design framework and hence, does not incur any overhead for address decoding.

In summary, the following contributions are made in this paper. \renewcommand{\labelenumii}{\Roman{enumii}}
\theenumiii
 \begin{enumerate} [(a)]
   \item Investigate DRAM peripherals and their security implications;
    \item Propose a novel Trojans in DRAM that can leak the data by exploiting the column multiplexing and self-refresh properties;
     \item Propose Trojans in DRAM that can tamper with refresh mechanism for polarity dependent fault injection;
     \item Demonstrate TrappeD exploits in action using an open-source RocketChip SoC system.

 \end{enumerate}

The paper is organized as follows: Section II reviews the basics of DRAM design and operation; Section III describes the proposed trojan trigger design and analysis; Section IV describes the DRAM vulnerabilities and the attack details; Section V describes the system implementation to validate the proposed Trojan attacks; Section VI presents discussions on the practicality, assumptions, and countermeasures to TrappeD; and finally Section VII draws the conclusion.

\section{Background}
\subsection{Basics of DRAM}

\begin{figure}[b]
        \centering
        \begin{subfigure}[b]{0.25\linewidth}
                \centering
              \includegraphics[width=0.99\linewidth]{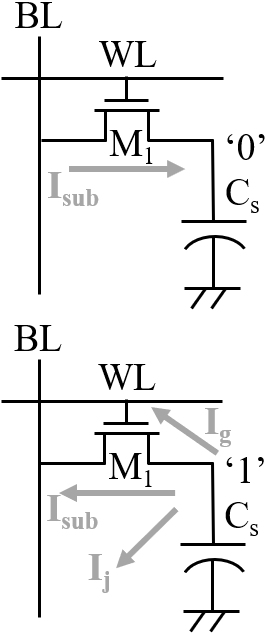}
              
              \vspace{-2mm}
                \caption{}
                \label{DRAM_bitcell}
                
        \end{subfigure}%
        \hspace{5mm}
        \begin{subfigure}[b]{0.5\linewidth}
                \centering
                \includegraphics[width=.9\textwidth]{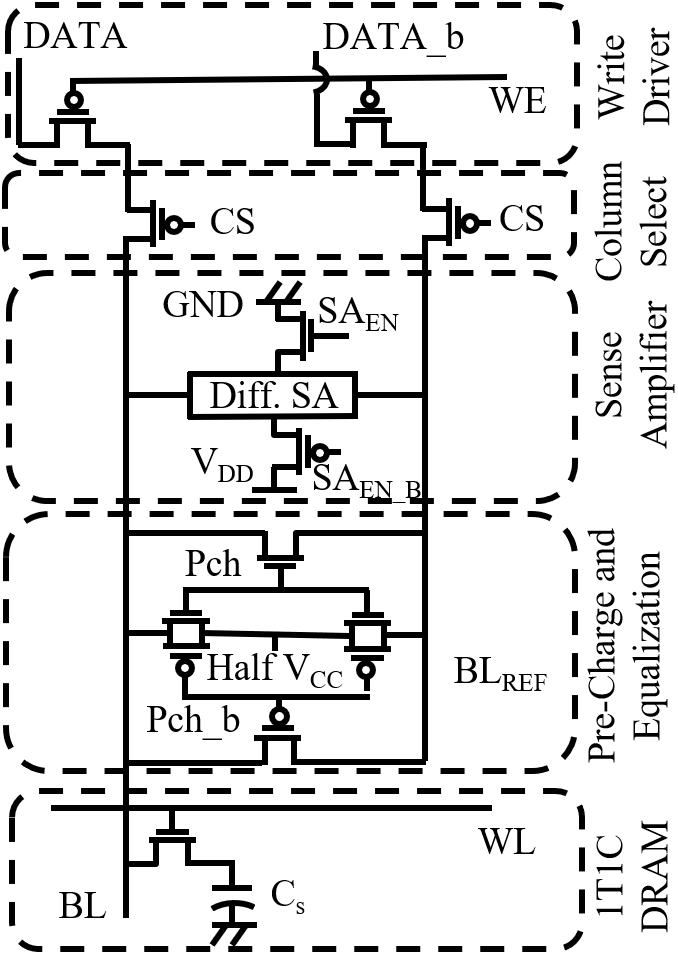}
                \caption{}
                \label{DRAM_arch}
        \end{subfigure}\\
        \vspace{-2mm}
        \caption{(a) 1T1C DRAM cell including leakage current sources for data `0' (up) and data `1' (down); (b) one DRAM column with peripherals.}
	\vspace{-2mm}
        \label{DRAM}
\end{figure}

Fig. \ref{DRAM_bitcell} shows a basic one-transistor, one-capacitor (1T1C) DRAM cell. The data is stored as charge in the capacitor $C\textsubscript{s}$. The charge corresponding to $V\textsubscript{dd}$ (0V) is considered as data `1' (`0'). The capacitor $C\textsubscript{s}$ gradually leaks or charges up over time and thereby, it requires to be read and written-back (i.e., refresh).     

\textbf{Architecture}: Fig. \ref{DRAM_arch} shows the column structure of an open bitline architecture \cite{jacob_ng_wang_2008}. It includes DRAM bitcell, $BL$, reference bitline ($BL\textsubscript{ref}$), sense amp, precharge circuit, column select, write driver and $Half\textunderscore V\textsubscript{cc}$ generator. 

The column circuit is implemented with the following features: a) Sense amp is placed on per-local-column basis to enable read/refresh of the selected column during read and refresh half-selected columns during write; b) Sense amp is activated by enabling both the header and footer transistors to prevent static current due to $Half\textunderscore V\textsubscript{cc}$ bitline precharge; c) The precharge and equalization circuit (equalizes $BL$ and $BL\textsubscript{ref}$ voltages) consists of full CMOS pass transistors; and d) $WL$ is boosted (to 1.6V in this paper) to write a full `1' through the NMOS access transistor and under-driven (to -0.2V in this paper) to reduce sub-threshold leakage during retention. Positive and negative charge pumps are used for $WL$ overdrive and underdrive respectively; and e) Column selection uses PMOS switch. 

\textbf{DRAM operation}: The DRAM operation can be classified into the following categories \cite{Ghosh_eDRAM}:

\textbf{Write:} During write, the access transistor $M\textsubscript{1}$ (Fig. \ref{DRAM_bitcell}) is turned ON by asserting the $WL$. The $BL$ is driven to $V\textsubscript{dd}$ (0V) for writing data 1 (0). The storage capacitor $C\textsubscript{s}$ charges up (discharges) to $V\textsubscript{dd}$ (0V) based on $BL$ voltage. Writing data `0' is accomplished by first writing a good `1' on the $BL\textsubscript{ref}$ while the $BL$ stays close to (threshold voltage of PMOS transistor) and then firing the sense amp. Once the sense amp is ON, it pulls the $BL$ all the way to 0V and a good `0' is written to the bitcell.
    
\textbf{Read:} This begins with pre-charging the $BL$ and $BL\textsubscript{ref}$ to $V\textsubscript{dd}/2$ and then the $WL$ is turned ON. The charge stored in $C\textsubscript{s}$ charges (discharges) the $BL$ if the stored data is `1' (`0'). Therefore, $C\textsubscript{s}$ also develops a differential and loses its stored value (destructive read). The resulting voltage differential in the $BL$ is converted to a digital value by the sense amp that compares the $BL$ to the reference voltage $BL\textsubscript{ref}$. 
    
\textbf{Write-back:} Following the $BL$ differential development, the sense amp is enabled. The $BL$ resolves to $V\textsubscript{dd}$ (0V) and $BL\textsubscript{ref}$ resolves to 0V ($V\textsubscript{dd}$) if the read value is `1' (`0'). The bitcell is restored to original value since the $WL$ is still ON after $BL$ resolves to the read value. 
    
\textbf{Stand-by (Retention):} A data `1' would lose its value once the charge reduces below $V\textsubscript{dd}/2$. Note that this very optimistic assumption. In reality, the sense amp requires $\sim$70mV of Sense Margin (SM) which puts maximum leakage of data `1' to  $\sim$0.640V. Three major sources of leakage are sub-threshold ($I\textsubscript{sub}$), junction ($I\textsubscript{j}$), and gate leakage ($I\textsubscript{g}$) through the access transistor (Fig. \ref{DRAM_bitcell}). Data `0' can lose its value if it charges above $\sim$0.360V (to guarantee 70mV of SM). The only source of leakage for data `0' is $I\textsubscript{sub}$. In our model, the charge stored in $C\textsubscript{s}$ only rises up to 30mV after which the rate of charging due to $I\textsubscript{sub}$ equals the rate of capacitor discharge. Therefore, data `0' does not suffer from retention issues.

\begin{figure}[t]
\vspace{-0mm}
        \centering  
        \begin{subfigure}[b]{0.42\linewidth}
                \centering
                \includegraphics[width=0.99\linewidth]{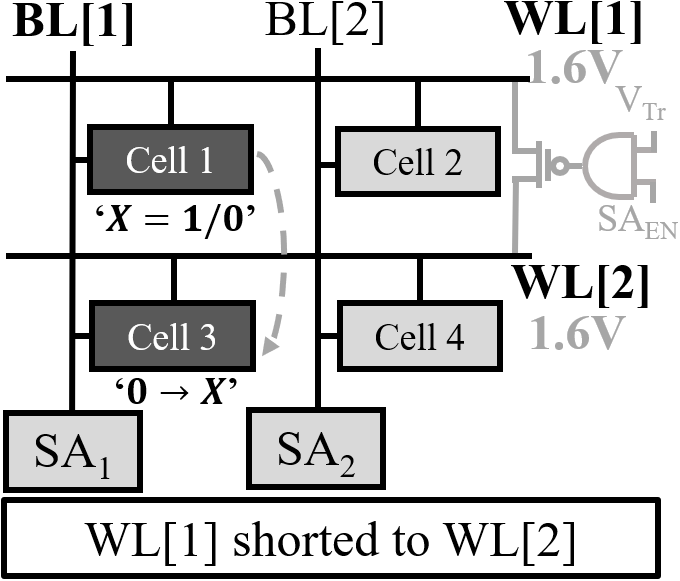}
                 \caption{}
                \label{WL_short_cir}
        \end{subfigure}%
        \hspace{4mm}  
        \begin{subfigure}[b]{0.4\linewidth}
                \centering
                \includegraphics[width=0.99\linewidth]{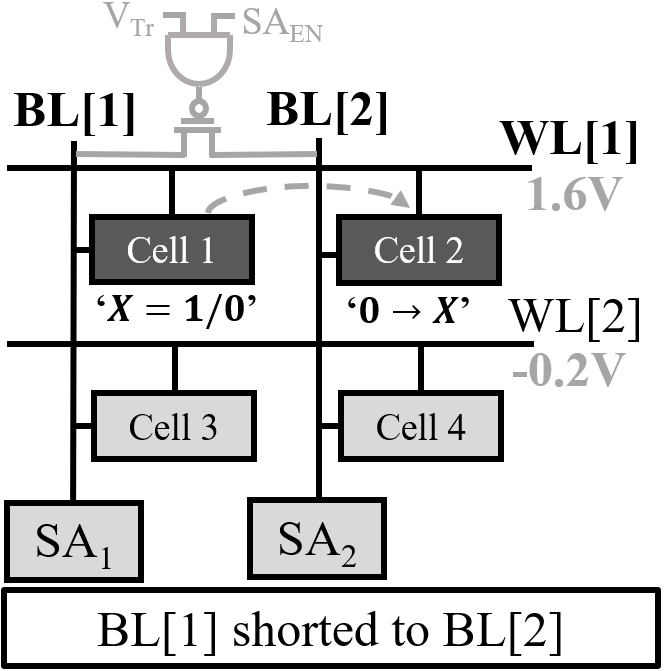}
                 \caption{}
                \label{BL_short_cir}
        \end{subfigure}\\
        \vspace{-2mm}
        \caption{DRAM data copy using a) $WL$ shorting; b) $BL$ shorting.}
	\vspace{-6mm}
        \label{Data_Copy}
\end{figure}

\section{Trojan Trigger Design}
\textbf{Design \cite{DATE_Nasim_2019}:} The trigger circuit (Fig. \ref{Trigger_DATE}) \cite{DATE_Nasim_2019} is designed to be activated if a particular memory address (chosen during design phase, let's call it $Add_{SET}$) is accessed for at least $N_{SET}$ times. The trigger has two inputs namely, $V(Add\textsubscript{SET})$ and $V(P\textsubscript{SET})$. $V(Add\textsubscript{SET})$ (= 1V in this work) is the wordline enable signal of $Add_{SET}$ and $V(P\textsubscript{SET})$ is a constant voltage source of 1V. For a more complex Trojan with a superior stealthiness, $V(P\textsubscript{SET})$ can be programmable (refer to Section \ref{Trojan_Design} for details).


\begin{figure*} [!t] 
        \centering
        
        \begin{subfigure}[b]{0.29\textwidth}
                \centering
                \includegraphics[width=0.8\linewidth]{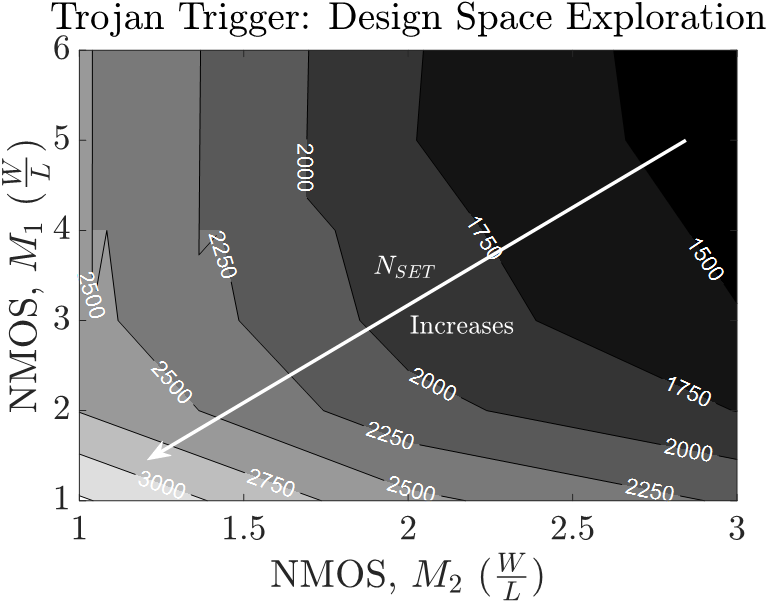}
                \vspace{-2mm}
                \caption{}
                \label{design_space_trigger_mode}
        \end{subfigure}%
        \hspace{0mm} 
        \begin{subfigure}[b]{0.3\textwidth}
                \centering
                \includegraphics[width=0.8\linewidth]{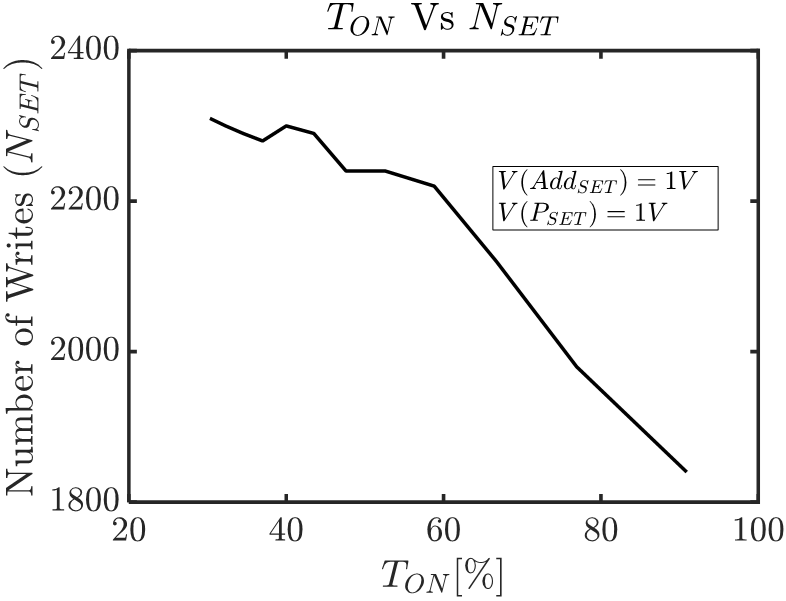}
                \vspace{-2mm}
                \caption{}
                \label{Ton_vs_write}
        \end{subfigure}%
        \hspace{0mm} 
        \begin{subfigure}[b]{0.3\textwidth}
                \centering
                \includegraphics[width=0.8\linewidth]{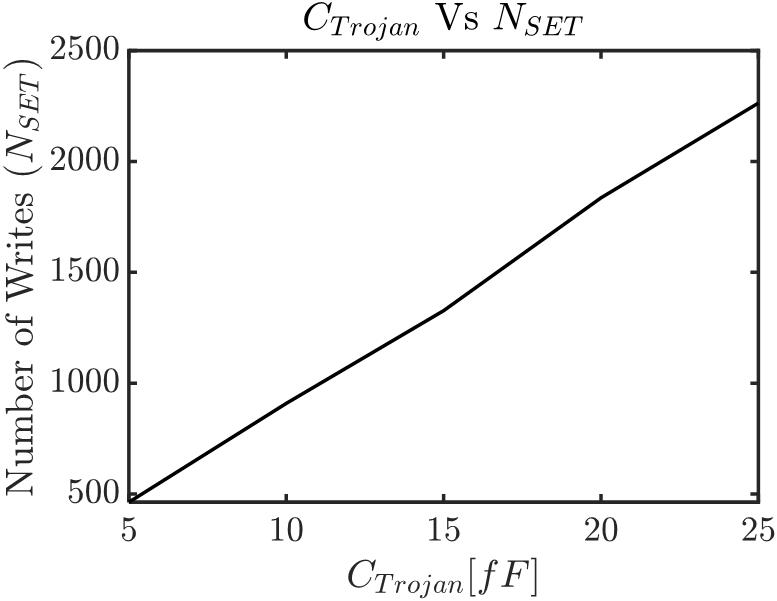}
                \vspace{-2mm}
                \caption{}
                \label{Ctrigger_vs_Write}
        \end{subfigure}
        \vspace{-1mm}
        \caption{(a) Design space exploration of Trojan trigger (Fig. \ref{Trigger_DATE}) with respect to (W/L) ratio of MOSFET $M_1$ and $M_2$; (b) required number of access ($N_{SET}$) increases if the adversary cannot access continuously; (c) a scaled down $C_{Trigger}$ leads to less number of $N_{SET}$ to activate the trigger. \cite{DATE_Nasim_2019}}
        \label{Nsets_ton_ctrigger}
        \vspace{-2mm}
\end{figure*}


Whenever $Add_{SET}$ is accessed, $V(Add\textsubscript{SET})$ is asserted and MOSFETs $M_{1}$ and $M_{3}$ are activated. $M_{2}$ has a thinner gate oxide compared to other MOSFETs and its source and drain are shorted. Therefore, $M_2$ works as a capacitor and charges $C\textsubscript{Trojan}$ from the $P\textsubscript{SET}$ source through Fowler Nordheim (FN) tunneling \cite{Ravindra_tunneling} if $V(Add\textsubscript{SET})$ is asserted. $M_{4}$ is an OFF transistor which offsets gate leakage of $M_{5}$ and prevents unwanted charging-up of node $X_{2}$. $M_{7}$ keeps node $X_3$ as low as possible until node $X_{2}$ charges up sufficiently. The node $X_4$, that is charged up during the hammering process, is used as the SET input of a SR latch. The output of the SR latch ($V_{Trigger}$) transitions from $0\rightarrow1$ when $X_4$ charges up to 0.5V. The signal $V_{Trigger}$ is then used to activate the Trojan. 

The charge at node $X_{2}$ leaks away (due to capacitance leakage of $C_{Trojan}$), once the hammering is discontinued. However, $V_{Trigger}$ will still be asserted due to the SR latch. In order to deactivate the Trojan, $V_{RESET}$ needs to be asserted. $V_{RESET}$ can be generated by accessing a different address (let's say $Add_{RESET}$) for at least $N_{RESET}$ times and using a circuit similar to the trigger one. Note that a smaller $C_{Trojan}$ ($\sim$1fF) can be used in the RESET circuit to minimize the area overhead which leads to $N\textsubscript{RESET} = $ 92. However, the AND'ed output of $V(Add_{RESET})$ and $V(P_{RESET})$ can also serve as $V_{RESET}$ which further reduces the area overhead.

\textbf{Simulation results:} Node $X_{2}$ charges up to 125mV (steady state) from all the leakage considering $V(P\textsubscript{SET})$ = 1V (worst-case charging due to leakage). This value is not enough to trigger the circuit. For the rest of the simulation, we have considered that $V(Add\textsubscript{SET})$ is a pulse source with ON/OFF time of 10ns/1ns. We consider the circuit to be triggered when $V\textsubscript{Trigger}$ reaches up to 0.5V. We started our analysis with $C_{Trojan}$ = 20fF.

Fig. \ref{design_space_trigger_mode} shows the design space exploration of trigger circuit considering two variables, the (W/L) ratios of MOSFETs $M_{1}$ and $M_{2}$. For a lower (W/L) ratio for both the MOSFETs, $N_{SET}$ increases. We have chosen (W/L) of $M_{1}$ and $M_{2}$ as 4 and 2 respectively for a sufficiently higher $N_{SET}$.


Next, we considered that the adversary accesses the preselected address for $T\textsubscript{ON}$ = 10$ns$ and then stays idle for $T\textsubscript{OFF}$ = 1/3/.../21/23 $ns$ and repeats this cycle. We found that the $C_{Trojan}$ does not significantly leak in the OFF cycle and the circuit can still be triggered with a higher $N_{SET}$ (Fig. \ref{Ton_vs_write}). We observe that the circuit will trigger even with a low $T_{ON}$ of 30$\%$. This means that it becomes harder to prevent Trojan activation using system level techniques such as limiting repeated access to one address.

Note that the attack gets auto reset without the SR latch since the node $X_{2}$ discharges (due to charge leak of $C_{Trigger}$) and eventually node $X_{4}$ goes down once the adversary stops the hammering after the trigger activates. Results indicate that the attack (charge at node $X_4$) lasts for 163.73$\mu$s if $Add_{SET}$ is accessed for 18$\mu$s. However, by adding the SR latch, the attack can last indefinitely until $Add_{SET}$ access is discontinued and $V_{RESET}$ is asserted.

A small $C_{Trigger}$ will require a low $N_{SET}$ to get the trigger activated. For example, the required number of access, $N_{SET}$ = 464 for $C_{Trigger}$ = 5fF  (Fig. \ref{Ctrigger_vs_Write}). This is still significantly high enough to evade the test phase. The value of $C_{Trigger}$ is chosen as 20fF since it offers a high $N_{SET}$ (= 1837) under nominal conditions, successfully triggers under all process corners and temperatures (-10\degree C to 90\degree C) and minimum $N_{SET}$ in the worst-case (= 68) is still high to evade testing phase.

Table \ref{tab3} summarizes the key performance metrics of the trigger. The absolute area and static power of proposed trigger are 42.95$\mu$m\textsuperscript{2} and 0.589$\mu$W, respectively which are 5.34$\times$10\textsuperscript{-5}\% and 6.24$\times$10\textsuperscript{-5}\% of a typical memory chip area and static power \cite{SRAM_area_power}, respectively. Therefore, the overhead due to the trigger is negligible to be detected via optical inspection or side channel analysis.

\begin{table}[h]
\vspace{-2mm}
\caption{Features of the Trojan Trigger}
\vspace{-3mm}
\begin{center}
\begin{tabular}{|c|c|}
\hline
	\textbf{Parameter} & \textbf{Trojan Trigger}\\ \hline\hline
	Dynamic Power ($\mu$W) & 3.781\\ \hline
	Static Power ($\mu$W) & 0.589 \\ \hline
	Energy/hammer (nJ) & 2.059 \\ \hline
    Area ($\mu$m\textsuperscript{2}) & 42.97 \\ \hline
    Target $N_{SET}$ & 1837 \\ \hline
    Worst-Case $N_{SET}$ & 1502\\ \hline
\end{tabular}
\vspace{-3mm}
\label{tab3}
\end{center}
\vspace{-1mm}
\end{table}

\begin{figure}[t]
        \centering
        \hspace{-1mm}
        \begin{subfigure}[b]{0.51\linewidth}
                \centering
                \includegraphics[width=0.99\linewidth]{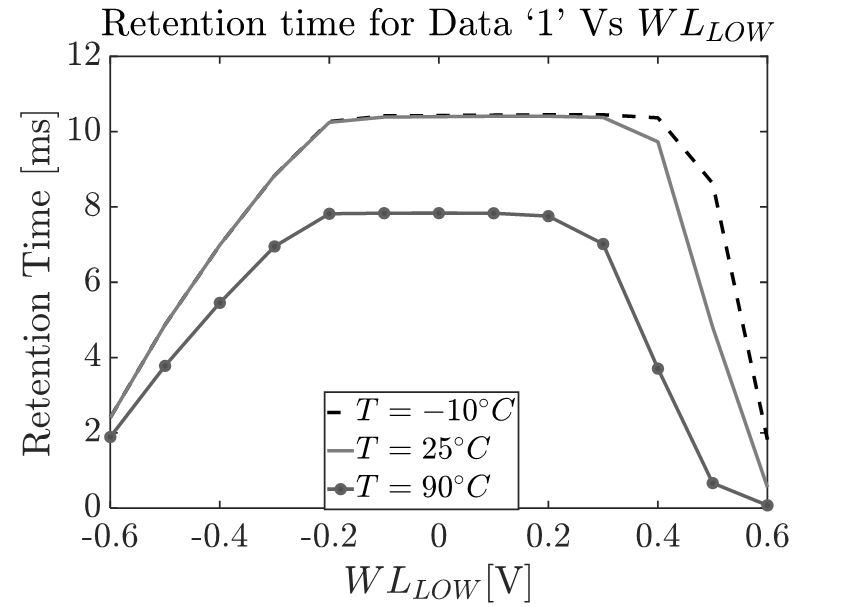}
                \caption{}
                \label{WL_Change}
        \end{subfigure}%
        \hspace{-1mm}
        \begin{subfigure}[b]{0.47\linewidth}
                \centering
                \includegraphics[width=0.99\linewidth]{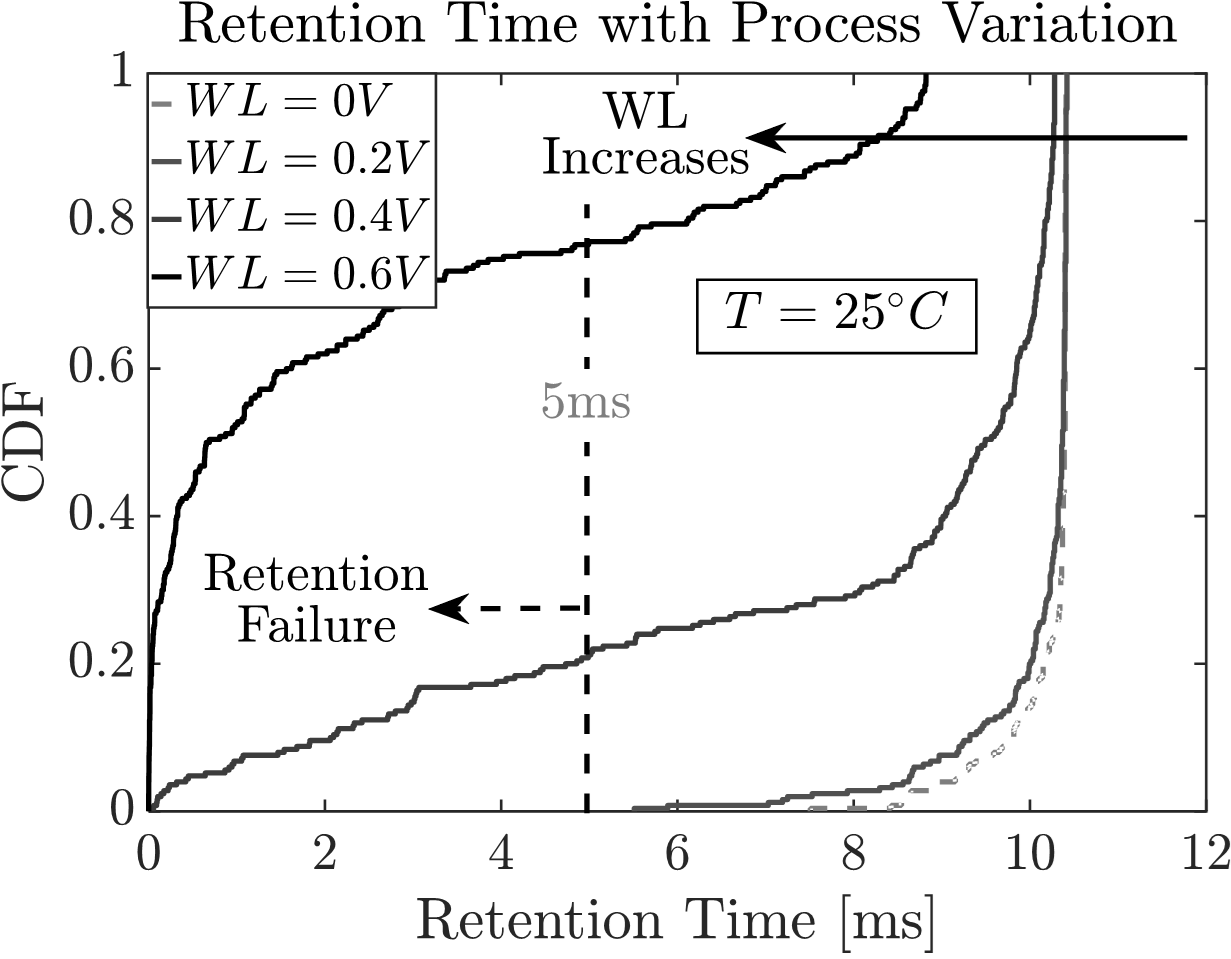}
                \caption{}
                \label{DRAM_retention}
        \end{subfigure}\\
        \vspace{-2mm}
        \caption{(a) DRAM retention time Vs $WL$ voltage during stand-by mode at different temperature; (b) retention failure increases considering  process variation at increased $WL$ voltage during retention.}
	\vspace{-2mm}
        \label{DRAM_retention_two_figures}
\end{figure}

\begin{figure}[t]
        \centering
        \hspace{-1mm}
        \begin{subfigure}[b]{0.27\linewidth}
                \centering
                \includegraphics[width=0.99\linewidth]{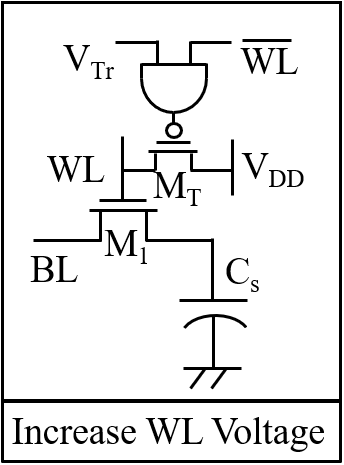}
                \caption{}
                \label{WL_overdrive}
        \end{subfigure}%
        \hspace{3mm}
        \begin{subfigure}[b]{0.49\linewidth}
                \centering
                \includegraphics[width=0.99\linewidth]{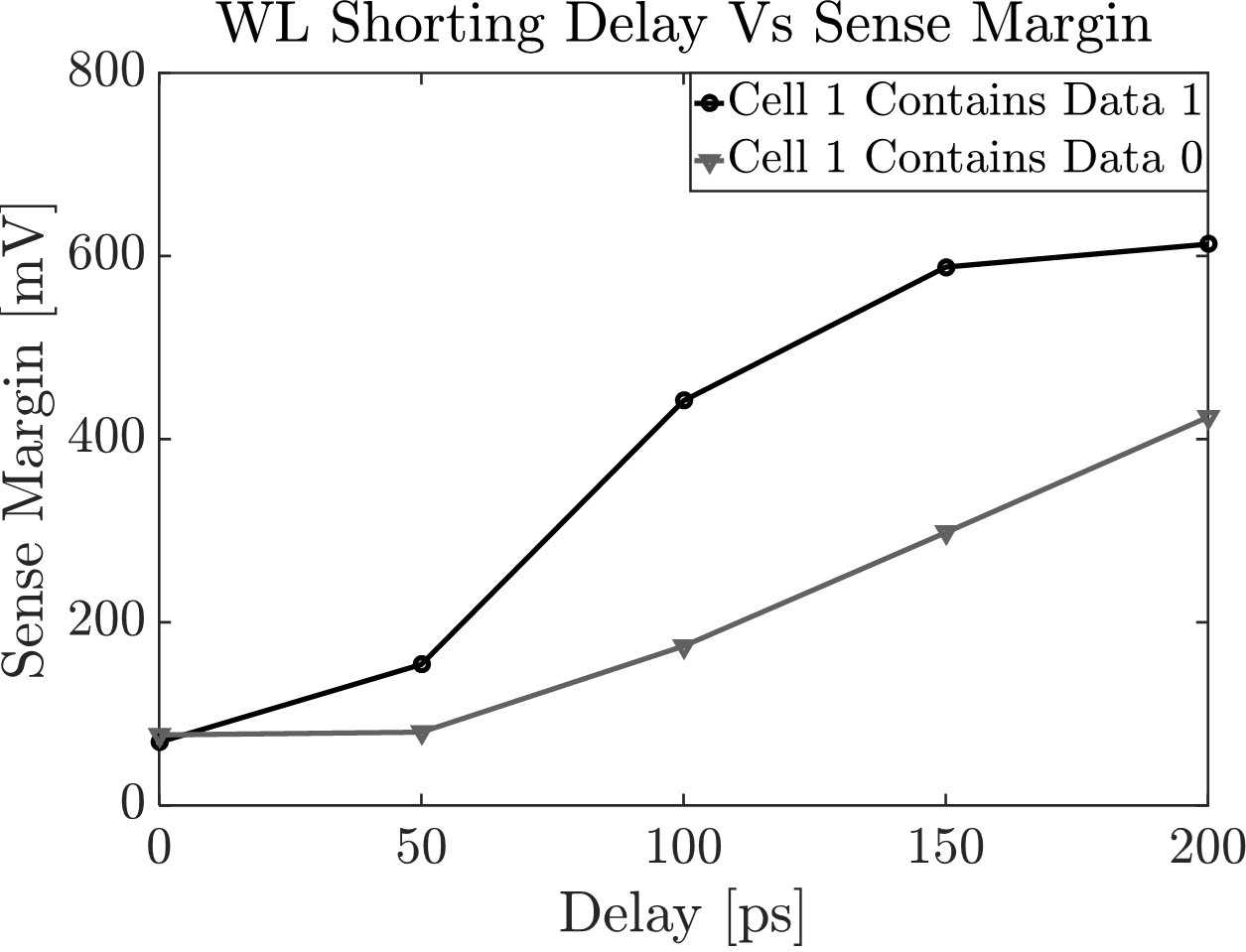}
                \caption{}
               \label{DRAM_WL_shorting_delay}
        \end{subfigure}\\
        \vspace{-1mm}
        \caption{(a) Method to increase $WL$ voltage; (b) $WL$ shorting delay (after the sense amp is fired) Vs sense margin for both data types stored in Cell 1.}
	\vspace{-5mm}
        \label{abnormal_two_figures}
\end{figure}

\section{DRAM Vulnerabilities and Trojans}
In this section, We present methods to exploit the vulnerabilities of the DRAM assist techniques and refresh mechanisms.

\begin{figure*} [!t] 
        \centering 
        \begin{subfigure}[b]{0.24\textwidth}
                \centering
                \includegraphics[width=0.99\linewidth]{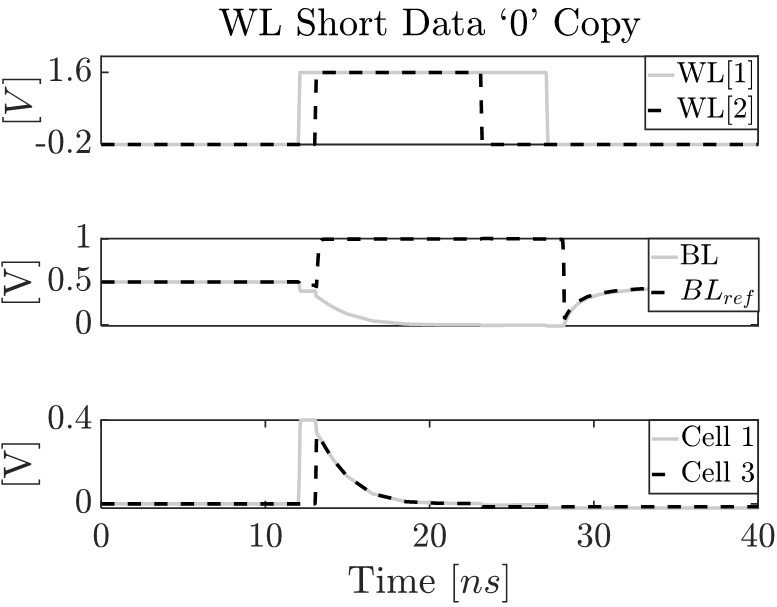}
                \vspace{-6mm}
                \caption{}
                \label{WL_0_copy}
        \end{subfigure}%
        \hspace{0mm}
        \begin{subfigure}[b]{0.24\textwidth}
                \centering
                \includegraphics[width=0.99\linewidth]{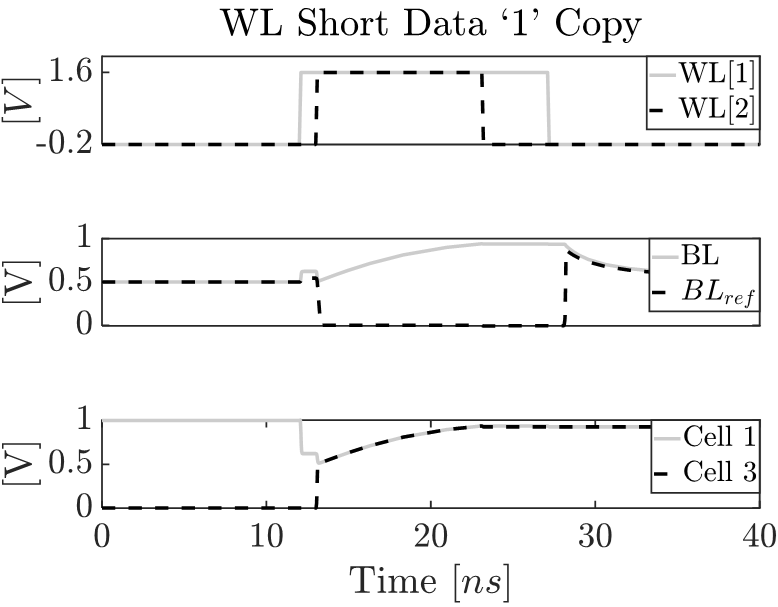}
                \vspace{-6mm}
               \caption{}
                \label{WL_1_copy}
        \end{subfigure}%
        \hspace{0mm} 
        \begin{subfigure}[b]{0.24\textwidth}
                \centering
               \includegraphics[width=0.99\linewidth]{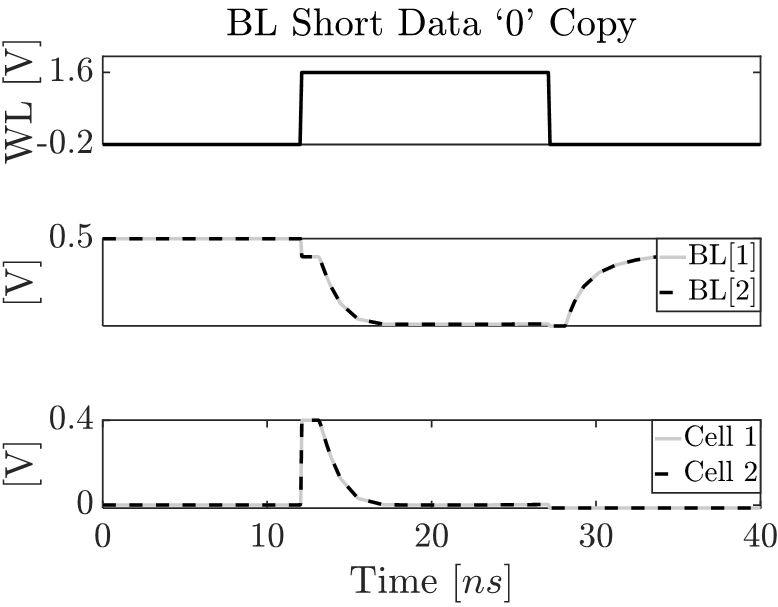}
               \vspace{-6mm}
                \caption{}
                \label{BL_0_copy}
        \end{subfigure}%
        \hspace{0mm}     
        \begin{subfigure}[b]{0.24\textwidth}
                \centering
                \includegraphics[width=0.99\linewidth]{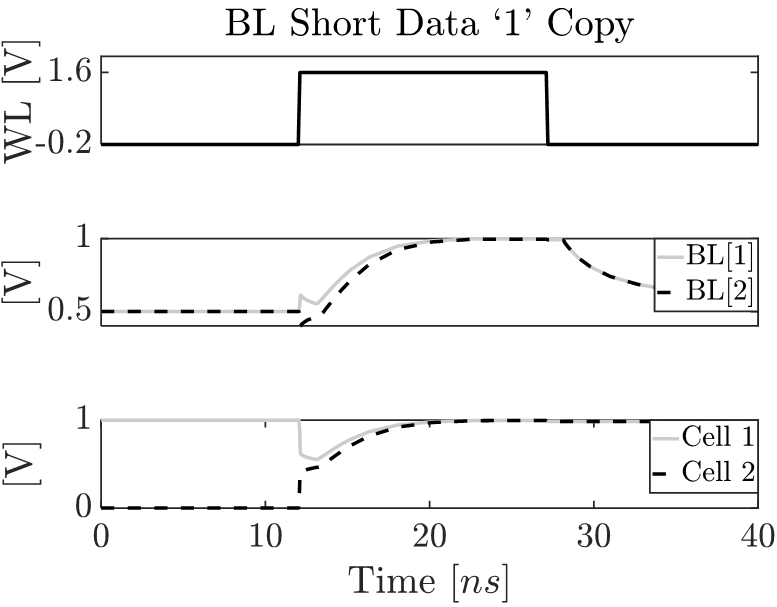}
                \vspace{-6mm}
               \caption{}
                \label{BL_1_copy}
        \end{subfigure}%
        \vspace{-1mm}
        \caption{Data leakage in DRAM through a) $WL$ shorting to copy data `0'; b) $WL$ shorting to copy data `1'; (c) $BL$ short to copy data `0'; (d) $BL$ shorting to copy data `1'.}
        \label{DRAM_Data_Leak}
        \vspace{-4mm}
\end{figure*}

\begin{figure}[t]

        \centering
        \begin{subfigure}[b]{0.49\linewidth}
                \centering
                \includegraphics[width=0.99\linewidth]{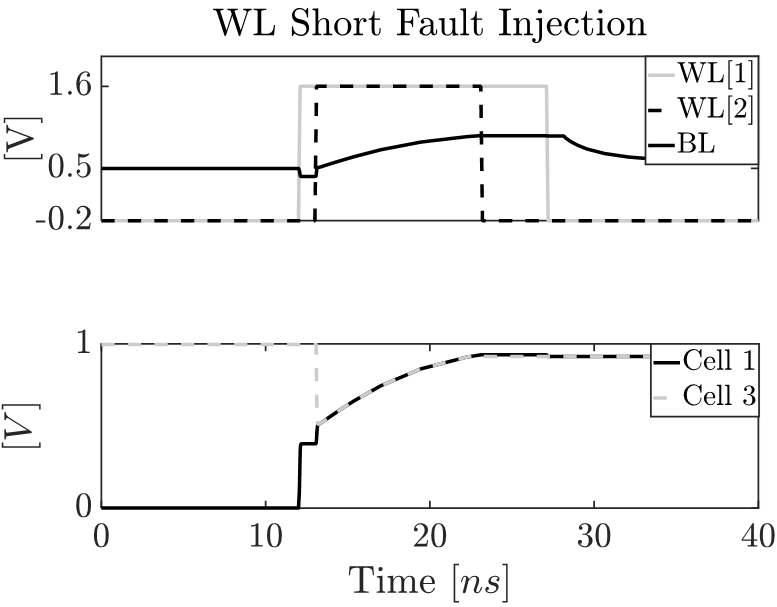}
                 \vspace{-2mm}
                \caption{}
                \label{WL_fault_inj}
        \end{subfigure} 
        \begin{subfigure}[b]{0.49\linewidth}
                \centering
                \includegraphics[width=0.99\linewidth]{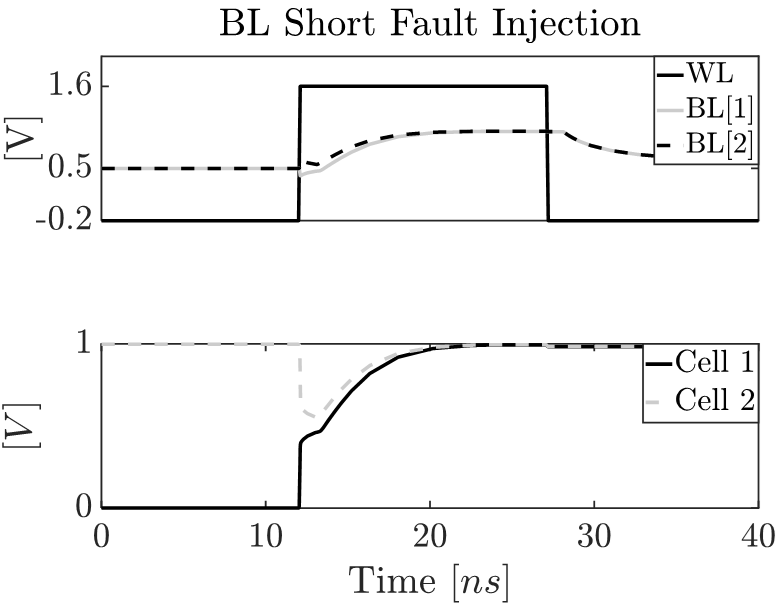}
                \vspace{-2mm}
                \caption{}
                \label{BL_fault_inj}
        \end{subfigure}
       	\vspace{-5mm}
        \caption{Fault Injection through (a) $WL$ shorting; (b) $BL$ shorting.}
	\vspace{-5mm}
        \label{DRAM_Fault_Injection}
\end{figure}

\subsection{DRAM assist and vulnerabilities}
\textbf{Wordline Underdrive and DoS}: 
Fig. \ref{WL_Change} shows that the retention time of data `1' changes as the $WL$ voltage during retention is swept. The worst case retention occurs at higher temperature. We consider retention failure if the capacitor discharges below $V\textsubscript{dd}/2$. Note that retention time decreases below -0.2V (due to Gate Induced Drain Leakage (GIDL)) and above 0.3V (due to subthreshold leakage) for each of the operating temperatures. The adversary cannot lower the $WL$ voltage below -0.2V since this is the lowest voltage available in the chip. However, adversary can increase the $WL$ voltage during retention. The $WL$ can be connected to $V\textsubscript{dd}$ through the transistor $M\textsubscript{T}$ (Fig. \ref{WL_overdrive}). The gate of $M\textsubscript{T}$ is controlled by the AND'ed signal of $V\textsubscript{Tr}$ (from Fig. \ref{Trigger_DATE}) and $\overline{WL}$ (Fig. \ref{WL_overdrive}). $\overline{WL}$ is used in order to ensure that the $WL$ underdrive only occurs during retention.  

We have performed 1000 point Monte-Carlo simulation with the same setup as SRAM to investigate the impact of increased $WL$ voltage during retention. We consider retention time below 5ms as failure. From Fig. \ref{DRAM_retention}, we note that if the adversary increases $WL$ voltage to 0.3V and 0.4V respectively from -0.2V during retention, the corresponding retention failure is 4.8\% and 43.6\% respectively, causing polarity fault injection attack.

\textbf{Information Leakage - $WL$ Shorting and Refresh:} Fig. \ref{WL_short_cir} shows the method to short $WL[1]$ and $WL[2]$. It is assumed that victim and adversary have control over Cell 1 and Cell 3, respectively. The shorting transistor is controlled  by the AND'ed signal of $V\textsubscript{Tr}$ and $SA\textsubscript{EN}$. $SA\textsubscript{EN}$ is used to ensure that the $WLs$ are shorted only after the sense amp is fired. This delay gives the sense amp sufficient differential between $BL$ and $BL_{ref}$ and ensures Cell 3 value does not corrupt Cell 1 (during sensing) that is being copied. In the AND Trojan, we get a delay of 40ps which provides a SM of 69mV. For larger SM an inverter chain can be used. Fig. \ref{DRAM_WL_shorting_delay} shows that SM for both data type increases with delay. Note that $WL[2]$ driver should be disabled during the shorting period.

When $WL[1]$ is asserted to 1.6V during a read operation of Cell 2, $WL[2]$ is also pulled to 1.6V after the sense amp firing. Since DRAM read operation is destructive and needs a write-back at the end of read operation, $BL[1]$ will resolve according to the stored value in Cell 1. Therefore, both Cell 1 and Cell 3 will get written to the previously stored value in Cell 1 since their $WLs$ are asserted. This will effectively copy the data to Cell 3.

Fig. \ref{WL_0_copy} shows when Cell 3 is initialized to `0' and data in Cell 1 is read as `0'. Both Cell 1 (due to write-back) and Cell 3 are written to `0'. Similarly, Fig \ref{WL_1_copy} shows that data in Cell 1 is read as `1' and both Cell 1 and Cell 3 are written to `1'. This data leak also applies when Cell 3 is initialized to `1'. If $WL$ shorting occurs with no delay after the sense amp activation, both Cell 1 and Cell 3 are written to Cell 3 data irrespective of the original Cell 1 data as shown in Fig. \ref{WL_fault_inj}. This can be done to inject a polarity dependent fault to Cell 1. Furthermore, the information leakage attack also works if write operation is performed to Cell 1 since the  $BL[1]$ will be driven to $V\textsubscript{dd}$ or 0V based on the write data and both $WL[1]$ and $WL[2]$ are shorted. One important point to note that all the cells sharing $WL[2]$ will be corrupted in this process.

\textbf{Information Leakage - $BL$ Shorting and Refresh:} $BL$ shorting is carried out in a similar way to $WL$ shorting (Fig. \ref{BL_short_cir}). It is assumed that victim and adversary have control over Cell 1 and Cell 2, respectively. The explanation for $WL$ shorting holds true for $BL$ shorting except none of the cells will be corrupted (since all the cells sharing  $WL[1]$ will be refreshed). Fig. \ref{BL_0_copy} shows that the data in Cell 1 is read as `0' and both Cell 1 (due to write-back) and Cell 2 (due to $BL[2]$ shorting) are written to `0'. Fig. \ref{BL_1_copy} shows the copy of data `1'. Similar to $WL$ shorting, $BL$ shorting can also be leveraged to inject a polarity dependent fault in Cell 1 as shown in Fig. \ref{BL_fault_inj}.  We restrict our discussion on this for the sake of brevity.

\subsection{Trojan design space exploration} \label{Trojan_Design}
We choose the capacitor-based trigger proposed in \cite{DATE_Nasim_2019} to generate the trigger signal ($V_{Tr}$) required to induce the retention failures (Fig. \ref{WL_overdrive}) and information leakage (Fig. \ref{Data_Copy}) attacks. This is because the design allows easy change of trigger parameters to vary the number of address accesses (denoted by $En_{Add}$) required to assert $V_{Tr}$. We modify the design parameters shown in Fig. \ref{Trigger_DATE} to allow the trigger design to stay undetected at all fast and slow corner cases and work under multiple temperatures (-10$\degree$C,25$\degree$C, and 90$\degree$C). The modified design is shown in Fig. \ref{Trigger_DATE}. Additionally, we include an SR latch to capture the trigger output.Note that the attack gets auto reset without the SR latch since the node $X_{2}$ discharges (due to charge leak of $C_{Trojan}$) and eventually node $X_{4}$ discharges once the adversary stops the hammering after the trigger activates. Results indicate that the attack (charge at node $X_4$) lasts for 163.73$\mu$s if $Add_{SET}$ is written for 18$\mu$s with $V(Add\textsubscript{SET})$ = 1V, $V(P\textsubscript{SET})$ = 1V. However, by adding the SR latch, the attack can last indefinitely until $Add_{SET}$ access is discontinued and $V_{RESET}$ is asserted.


\section{System Exploits using TrappeD}

\subsection{Overview of systems architecture}
For the purposes of demonstration, we show a simple information leakage attack using TrappeD. We use a RISC-V based RocketChip \cite{rocketchip} SoC template for implementing the TrappeD system. Our RocketChip SoC is configured with an in-order Rocket Core, a fast L1 data cache, and connects to an external DRAM on the AXI port (mapped to \texttt{0x80000000}), which serves as the main memory for the system.

\subsection{TrappeD trojan deployment}
\begin{figure} [t]
\vspace{-0mm}
 \begin{center}
    \includegraphics[trim=0 0 0 0,clip,width=0.49\textwidth]{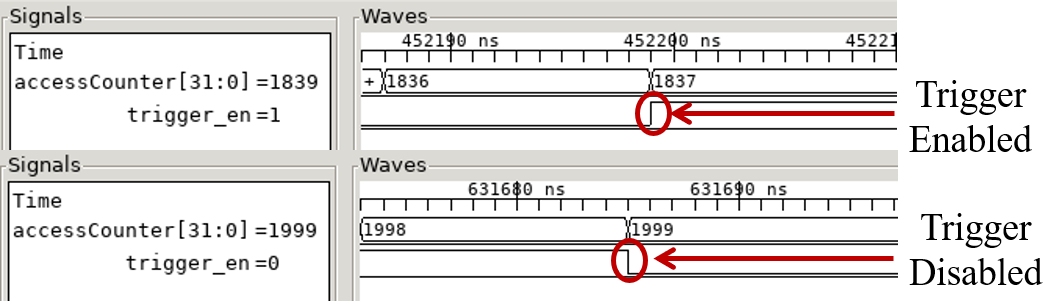}
 \end{center}
 \vspace{-3mm}
 \caption{RocketChip emulator waveform showing \texttt{Trigger\_en} signal enabled after 1837 \texttt{adversary} accesses and disabled after 1999 accesses.} \label{trigger_wave}
 \vspace{-4mm}
\end{figure}

We modified the RocketChip code to inject the Trapped trigger and payload. The trigger logic is introduced in the D-cache and the payload logic in the DRAM. For modelling the effects of trigger, we instantiate a 32-bit register (\texttt{AccessCounter}) that is initialized to `0'. Accesses to the \texttt{trigger} address (= \texttt{0x80022328}) increments the register value by `1'. We introduce logic to detect if these increments have reached our $N_{SET}$ threshold of 1837, to active the trigger signal (\texttt{trigger\_en}). The trigger signal is sent to the payload logic. Once triggered, the signal stays asserted to facilitate the attack till the adversary accesses the \texttt{trigger} address an additional 163 times. The waveform generated from the RocketChip hardware emulation depicting \texttt{trigger\_en} enabled and disabled is shown in Fig. {\ref{trigger_wave}}. The payload activates when \texttt{trigger\_en} is high. To model the effects of the payload, we select two addresses: (i) an adversary controlled address, \texttt{adversary} (= \texttt{0x800222E8}), and (ii) a victim address, \texttt{victim} (= \texttt{0x800222A8}). The payload logic shorts the two addresses, effectively copying the data from \texttt{victim} to \texttt{adversary}, facilitating an information leakage attack.

\subsection{Attack demonstration}
\begin{figure} [t!]
\vspace{-0mm}
 \begin{center}
    \includegraphics[trim=0 0.23in 0.8in 0.5in,clip,width=0.4\textwidth]{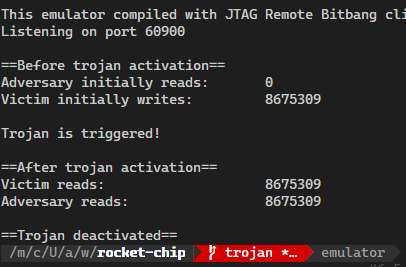}
 \end{center}
 \vspace{-1mm}
 \caption{RocketChip emulator console output for the attack code in Listing 1. It can be seen that $adversary$ reads \texttt{`8675309'} after Trojan activation.} \label{codeoutput}
 \vspace{-4mm}
\end{figure}

We demonstrate the attack with the help of a simple C program as shown in Listing {1}. The code is compiled with the \texttt{riscv-gcc} compiler to generate a RISC-V ELF binary that can be run on the cycle-accurate emulator for the RocketChip system. For simplicity, we have assumed that the \texttt{adversary} and \texttt{victim} addresses are part of the same process, and the adversary can access \texttt{adversary}, but not \texttt{victim}. We also assume that the adversary has control over the number of times the \texttt{trigger} address can be accessed. The program initializes contents of \texttt{adversary} to `0' and \texttt{victim} to the confidential information `8575309'. The program loops over and accesses \texttt{trigger} 1837 times, which is greater than the minimum accesses required by the Trojan trigger as set by $N_{SET}$, thus activating the \texttt{trigger\_en} signal. The payload causes an information leakage by copying the confidential contents of \texttt{victim} to \texttt{adversary}. The adversary can now read out from \texttt{adversary} address and thus reveal the confidential information `8675309', as shown in Fig. \ref{codeoutput}.


\begin{figure} [t]
\vspace{-0mm}
 \begin{center}
    \includegraphics[width=0.49\textwidth]{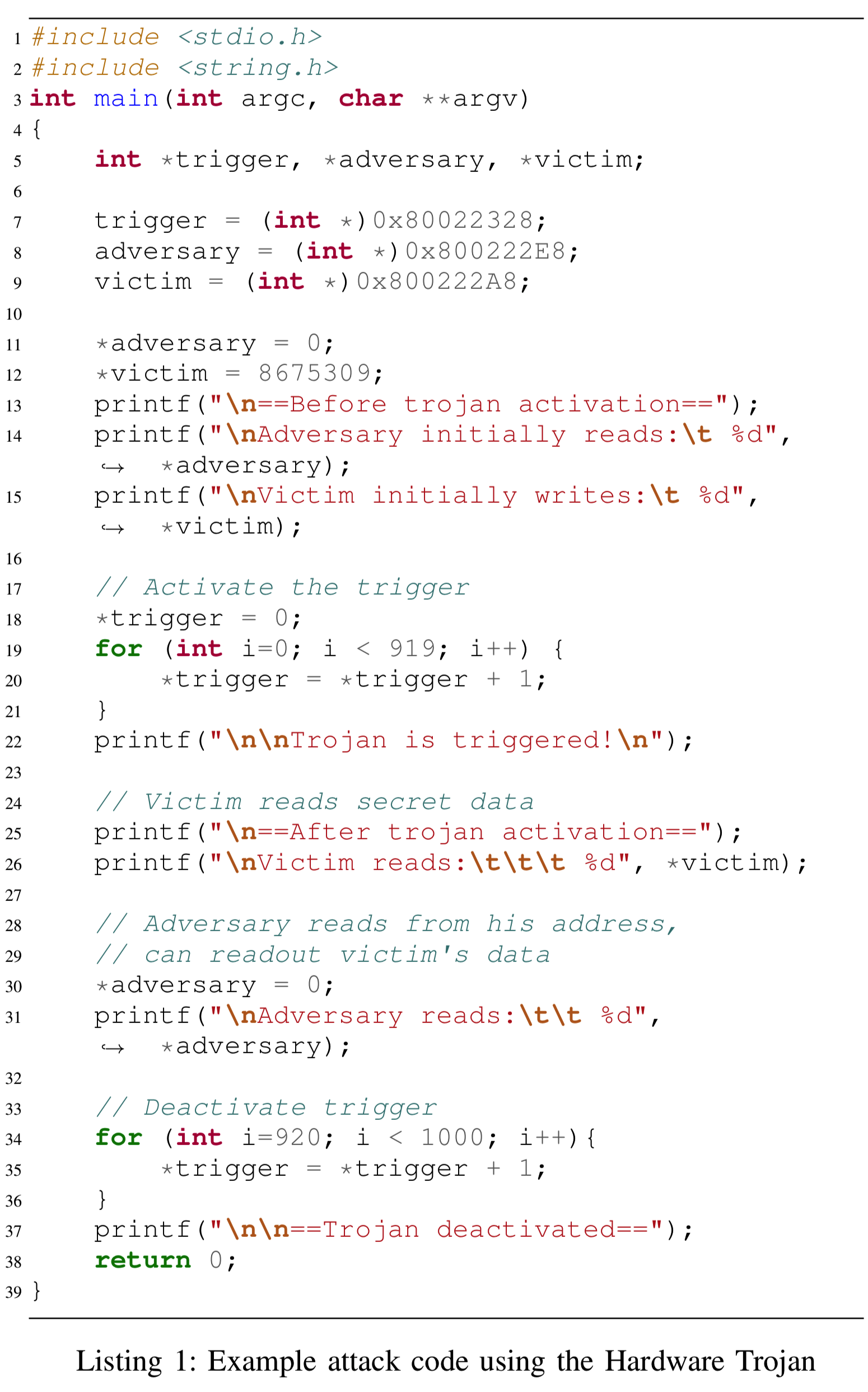}
 \end{center}
 \vspace{-3mm}
 \caption*{} 
 \vspace{-4mm}
\end{figure}

\section{Discussion}

\subsection{Fault injection and DoS Attack}
If read/write/retention failure occurs for one polarity (either for data `0' or `1'), it is considered as fault injection. Such attack can leak system assets \cite{Fault} such as, keys. One example is setting plaintext to all 0 or 1 by injecting faults during crypto operation which makes the ciphertext (that is sent out) same as keys, and can be recovered by the adversary. However, if failure occurs for both polarities, it is considered as DoS attack.

\subsection{Bypassing error detection}
In state-of-the-art memory, techniques like Cyclic Redundancy Check (CRC) or Error Correcting Code (ECC) \cite{ECC_CRC} is implemented. The ECC and/or CRC word is computed for the raw data and written along with data during write operation. During read operation, CRC/ECC is again calculated based-on the read data and matched with the stored CRC/ECC. If read or even write operation incurs an error, CRCs/ECCs will not match and the data can be discarded. Furthermore, ECC can correct 1/2 bits of error (based on the implementation). Therefore, fault injection will fail and adversary can only launch DoS attack. However, if the CRC or ECC bits are also tampered to match the data with injected fault, the manipulated data will be considered as valid data.

\subsection{Possible defenses}
i) Dummy Bits: Each row of the memory can be designed with few dummy bits. During run-time, each row can be written with known dummy bits and read to validate. During Trojan induced fault/DoS, the dummy bits will fail which can be detected as an attack. However, information leakage cannot be detected. Note that this will incur slight area and power overhead.

ii) Trusted ECC: {The current implementation of ECC adds a few global columns in the memory. For example, if the data width is 64 bit and ECC needs 5 bits, a total of 69 global columns are implemented in the memory array. This is a vulnerability since the ECC bits can also be tampered. We propose to separate the ECC bits from the data bits and store them in a trusted memory known to be Trojan free through rigorous validation (possible due to small size). This way the fault injection, DoS and information leakage attacks can be detected since ECC bits can be checked to detect the tampering at run-time (once Trojan is activated). }

\balance

\section{Conclusions}
{In this paper, we investigated the advanced circuit features employed in the peripherals of nanometer memories e.g., WLUD and wordline overdrive for DRAM from a security perspective. We show that an adversary can manipulate these features to launch fault injection attacks and DoS. The adversary can also leverage wordline shorting and bitline shorting in order to leak sensitive information. We also demonstrated the feasibility of launching system exploits leveraging TrappeD on RocketChip SoC platform.}

\vspace{2mm}

\textbf{Acknowledgements}: 
This work is supported by SRC (2847.001), NSF (CNS- 1814710, CNS- 1722557, CCF-1718474, DGE-1723687 and DGE-1821766) and DARPA Young Faculty Award (D15AP00089).

\addtolength{\textheight}{-12cm}   









\bibliographystyle{ieeetr}
\bibliography{main}

\end{document}